\documentclass[12pt,dvips,oneside]{article}
\usepackage{booktabs,graphicx,overcite,xspace,amsmath,amssymb,longtable,multirow,algorithm,algorithmic}
\usepackage[scanall]{psfrag}
\usepackage[small,sc]{caption}
\usepackage[dvips]{color}
\usepackage{color}
\usepackage{setspace}
\normalsize

\begin{document}

\title{Protein Structure Prediction Using Basin-Hopping}
\author{Michael C. Prentiss, \\
Center for Theoretical Biological Physics, University of California at San Diego, \\
La Jolla, CA 92093, USA \\
David J. Wales, \\
University Chemical Laboratories, Lensfield Road, Cambridge CB2 1EW, \\
United Kingdom \\
and Peter G. Wolynes \\
Center for Theoretical Biological Physics, University of California at San Diego, \\
La Jolla, CA 92093, USA}

\maketitle
\newpage
\vfill

\begin{abstract}
Associative memory Hamiltonian structure prediction potentials are not overly rugged, thereby
suggesting their landscapes are like those of actual proteins. In the present contribution we show
how basin-hopping global optimization can identify low-lying minima for the corresponding mildly
frustrated energy landscapes. For small systems the basin-hopping algorithm succeeds in locating
both lower minima and conformations closer to the experimental structure than does molecular
dynamics with simulated annealing. For large systems the efﬁciency of basin-hopping decreases for
our initial implementation, where the steps consist of random perturbations to the Cartesian
coordinates. We implemented umbrella sampling using basin-hopping to further conﬁrm when the
global minima are reached. We have also improved the energy surface by employing bioinformatic
techniques for reducing the roughness or variance of the energy surface. Finally, the basin-hopping
calculations have guided improvements in the excluded volume of the Hamiltonian, producing
better structures. These results suggest a novel and transferable optimization scheme for future
energy function development.
\vfill
\end{abstract}

\maketitle

\section{Introduction}
The complexity of the physical interactions that guides the folding of 
biomolecules presents a significant challenge for atomistic modeling.  
Many current protein models use a coarse grained approach to remove degrees 
of freedom, such as non-polar hydrogens, which increases the feasible 
time step in molecular dynamics simulations \cite{NAMD,GROMACS}. 
For a more dramatic improvement of the computational efficiency 
solvent degrees of freedom can be reduced \cite{levyonuchic}.  
In this case more severe 
approximations can prevent the model from reproducing experimental results. 
Another option is to reduce the number of degrees of freedom of the solute.
The associative memory Hamiltonian (AMH) \cite{FriedrichsMS89, 
FriedrichsMS90, FriedrichsMS91}, is a coarse-grained molecular mechanics 
potential inspired by 
physical models of the protein folding process, but flexibly 
incorporates bioinformatic data to predict protein structure.  
The AMH is first optimised using the minimal frustration principle
in terms of the $T_{f}/T_{g}$ ratio, which estimates the separation 
in energy relative to the variance for the misfolded ensemble.  Along with 
using the energy of the native structure to estimate $T_{f}$, a random energy
model\cite{DerridaB81} estimate of the glass transition temperature, 
$T_{g}$, is used based on a set of decoy structures. $T_{g}$ 
represents a characteristic temperature scale at which kinetic 
trapping in misfolded states dominates the dynamics.
An improved potential is next obtained that uses better estimates of 
the $T_{f}/T_{g}$ ratio obtained by maximizing the normalized 
difference between the native state and a sampled set of misfolded decoys 
which are self-consistently obtained from the potential itself.
The potential so obtained is transferable for the prediction of structures
outside of the training set.  The ratio $T_{f}/T_{g}$ 
has provided a powerful metric for the optimisation of 
both this bioinformatically informed energy function 
\cite{GoldsteinRA-AMH-92, Baker_2007},  as well as other types of 
energy functions incorporating only physical information
\cite{Scheraga_2001,takada_wolynes,takada_06}. 

While the optimisation \cite{opt_note} of parameters using a training set of evolved
natural proteins smooths the energy landscape from what it would be for a 
random hetero-polymer, the common problem of multiple competing minima 
persists even when using a reasonably accurate structure prediction 
potential, such as this one. 
Simulated annealing with molecular dynamics has previously been used to 
search the rugged landscapes of optimised structure prediction 
potentials \cite{Zong_jacs}.  While free energy profiles indicate that better 
structures actually are present at low temperatures, the slow kinetics of 
a glass-like transition during annealing has prevented these minima from being 
reached \cite{Eastwood00}. To quantitatively investigate the origin of the 
sampling difficulties it is desirable to use different search strategies.  

\begin{figure}[htb]
\psfrag{Energy}[cr][cr]{Energy}
\centerline{\includegraphics[width=0.3\textwidth,angle=-90]{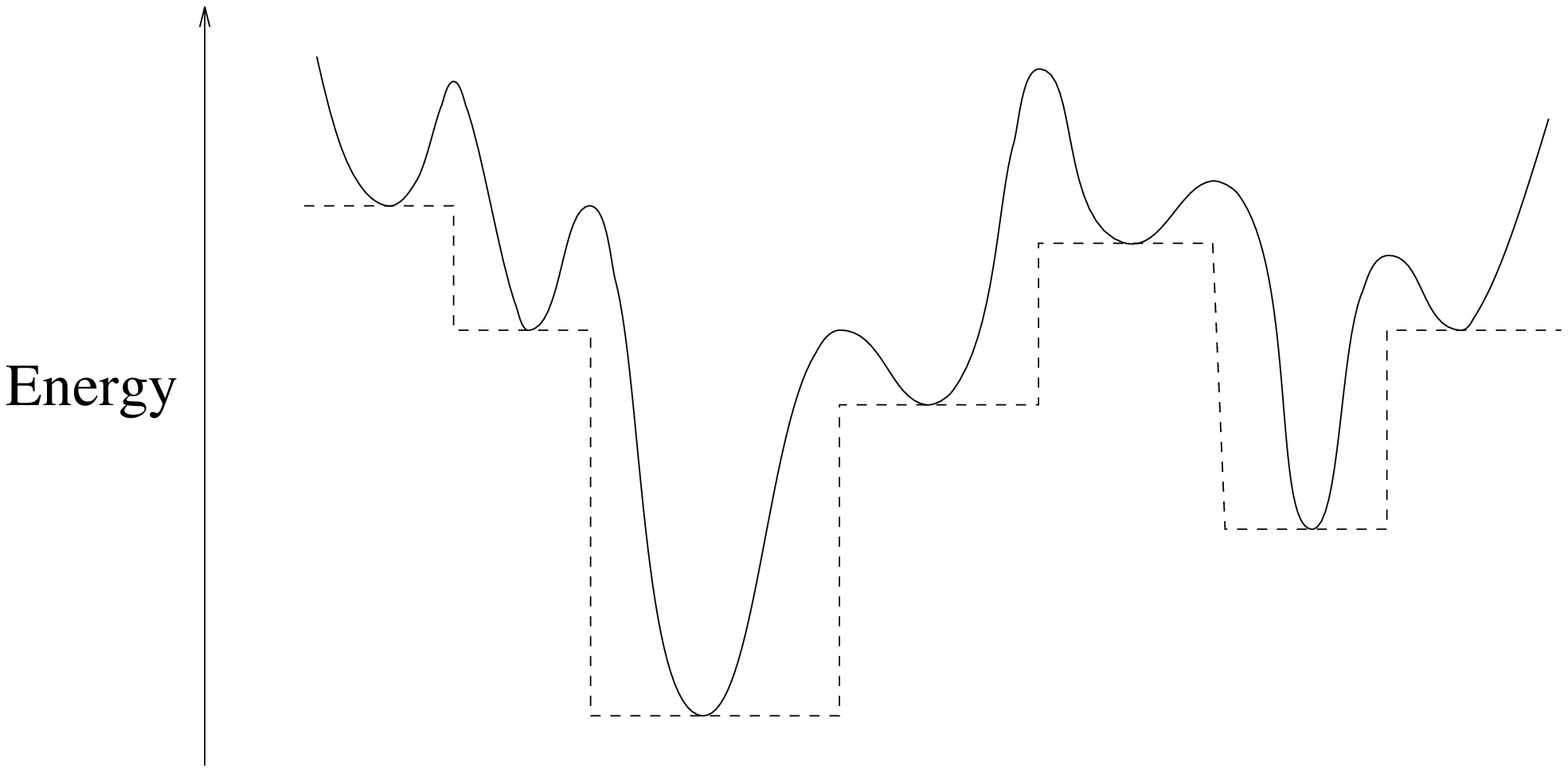}}
\caption{\label{transform}
}
\end{figure}

Here we implement the basin-hopping 
global optimisation algorithm \cite{WalesDJDoyeJPK_1997,lis87,waless99},
which has proved capable of overcoming large energetic barriers in a wide 
range of different systems.  Basin-hopping is an algorithm where a 
structural perturbation is followed by energy minimisation. This procedure effectively 
transforms the potential energy surface, by removing high barriers, as shown in 
Fig.~\ref{transform}. Moves between local minima are accepted or rejected 
based upon a Monte Carlo criterion. 
Avoiding barriers by employing a numerical minimisation step not only 
facilitates movement between local minima, but also broadens their 
occupation probability distributions, which overlap over a wider temperature 
range, thereby increasing the probability of interconversion \cite{doyew98b}. 
Furthermore, it does not alter the nature of the local minima 
since the Hamiltonian itself is not changed, enabling the  
comparison between molecular dynamics and basin-hopping generated minima.
This method has previously been applied to find global minima in atomic 
and molecular clusters \cite{Wales03,CCD}, biopolymers 
\cite{CarrW05,verma_wenzel}, and solids \cite{MiddletonHMW01}. 
Since the  algorithm only requires coordinates, energies, and gradients,
it can be transferred between different molecular systems such as
binary Lennard-Jones clusters, all-atom biomolecules, or coarse-grained 
proteins models as in this study.

\section{Theory and Computational Details}
The AMH energy function used in the present work has 
previously been optimised over a set of non-homologous 
$\alpha$ helical proteins, and consists of a backbone term, $E_{\rm back}$, and
an interaction term, $E_{\rm int}$, which has an additive form 
\cite{Hardin00,Eastwood02}.  This
model is sometimes termed the AMC model (associative memory contact) to 
distinguish it from one that uses nonadditive water mediated interactions 
termed the AMW model \cite{papoian_04,Zong_jacs}.
Since this model has been described in detail before \cite{Eastwood00,
Prentiss06}, we will only summarize its form here. We 
employ a version of the coarse-grained model where the twenty 
letter amino acid code has been reduced to four, and the
number of atoms per residue is limited to three (C$_\alpha$, C$_\beta$, 
and  O), except for glycine residues.  The units of energy and temperature 
were both defined during the parameter optimisation.
The interaction energy $\epsilon$  was defined in terms of the 
native state energy
excluding backbone contributions, $E^{\rm N}_{\rm amc}$, via
\begin{equation}
\label{units}
\epsilon=\frac{\left\vert E^{\rm N}_{\rm amc}\right\vert}{4N},
\end{equation}
where $N$ is the number of residues of the protein being considered.
Temperatures are quoted in terms of the reduced temperature 
${T_{amc}}=k_{\rm B}T/\epsilon$.  While \(E_{\rm back}\) creates 
self-avoiding peptide-like stereochemistry, \(E_{\rm int}\)
introduces the majority of the attractive interactions that produce
folding. The interactions described by $E_{\rm int}$ depend on the
sequence separation $\left\vert i-j \right\vert$.  The interaction
between residues less than 12 amino acids apart were defined 
by Eqs.~(\ref{amhsrmr}). 
\begin{equation}
 \label{amhsrmr}
E_{\rm local}=-\frac{\epsilon}{a}\sum_{\mu=1}^{N_{mem}}\sum_{j-12\le i\le j-3}
\gamma(P_i,P_j,P_{i'}^\mu,P_{j'}^\mu,x(\left\vert i-j\right\vert))
\exp\left[-\frac{(r_{ij}-r_{i'j'}^{\rm \mu})^2}{2\sigma^2_{ij}}\right],
\end{equation}
The index $\mu$ runs over all $N_{mem}$ memory proteins to which
the protein has previously been aligned using a sequence-structure threading algorithm
\cite{KoretkeKK96} ({\it i.e.} each $i$-$j$ pair in the protein 
has an $i'$-$j'$ pair associated with it in every
memory protein; if, due to gaps in the alignment, there is 
no $i'$-$j'$ pair associated with $i$-$j$
for a particular memory then this memory protein simply gives no 
contribution to the interaction between residues $i$ and $j$). The 
interaction between C$_\alpha$ and C$_\beta$ atoms is a sum of
Gaussian wells centred at the separations $r_{i'j'}^{\rm \mu}$ of 
the corresponding memory atoms.  The widths of the Gaussians are
given by $\sigma_{ij}=\left\vert i-j\right\vert^{0.15}$ \AA. The scaling 
factor $a$ is used to satisfy Eq.~\ref{units}.
The weights given to each well are controlled by
$\gamma(P_i,P_j,P_{i'}^\mu,P_{j'}^\mu,x(\left\vert i-j\right\vert))$, which
depends on the identities
$P_{i'}$ and $P_{j'}$ of the residues to which $i$ and $j$ are aligned, as well as the identities $P_i$ and $P_j$ of $i$ and $j$ themselves.
The self-consistent optimisation calculates the $\gamma$ parameter which originates the cooperative folding in the model.
A three well contact potential [Eq.~(\ref{amhlr})] is used for residues 
separated by more than 12 residues as described by,
\begin{equation}
 \label{amhlr}
E_{\rm contact}=-\frac{\epsilon}{a}
\sum_{i<j-12}\sum_{k=1}^3\gamma(P_i,P_j,k)c_k(N)U(r_{min}(k),r_{max}(k),r_{ij}).
\end{equation}
  
Here, the sequence indices $i$ and $j$ sum over all pairs of 
C$_\beta$ atoms separated 
by more than 12 residues. The sum over $k$ is over the three wells 
which are approximately  square wells between $r_{min}(k)$ and
$r_{max}(k)$ defined by,
\begin{equation}
U(r_{min}(k),r_{max}(k),r_{ij})=\frac{1}{4}\left\{\left[1+\tanh\left(7[r_{ij}-r_{min}(k)]/\text{\AA}\right)\right]
+\left[1+\tanh\left(7[r_{max}(k)-r_{ij}]/\text{\AA}\right)\right]\right\}.
\end{equation}
The parameters ($r_{min}(k),r_{max}(k)$), are (4.5\AA, 8.0\AA) , (8.0\AA, 10.0\AA), and (10.0\AA, 15.0\AA) for $k=1,2$ and 3 respectively. In order to 
approximately account for the variation of the probability distribution 
of pair distances with number of residues in the protein ($N$) a factor 
$c_k(N)$ has been included in $E_{\rm long}$. It is given by
$c_1=1.0$, $c_2=1.0/(0.0065N+0.87)$ and $c_3=1.0/(0.042N+0.13)$. The individual wells are also weighted by $\gamma$ parameters which depend on the 
identities of the amino acids involved, using the 4-letter code defined above.
In contrast to the interactions between residues closer in sequence, 
this part of the potential does not depend on 
the database structures that define local-in-sequence 
interactions.

To pinpoint the effects of frustration or favorable non-native contacts always
present in any coarse gained 
protein model, we simulated a perfectly smooth energy 
function often called a G\=o model \cite{go83}. 
G\=o models are an essential tool for understanding protein folding 
kinetics \cite{portman98,Koga_Takada01}. 
While having the same backbone terms \cite{Eastwood01}, in 
this single structure based Hamiltonian [Eq.~(\ref{go_eq})], 
all of the interactions, \(E_{\rm int}\) are 
defined by Gaussians whose minima are located at the pair distribution 
found in the experimental structure:
\begin{equation}
 \label{go_eq}
E_{\rm G{\bar o}}^{\rm AM}=-\frac{\epsilon}{a_{\rm G{\bar o}}}\sum_{i\le j-3}
\gamma_{\rm G{\bar o}}(x(\left\vert i-j\right\vert))
\exp\left[-\frac{(r_{ij}-r_{ij}^{\rm N})^2}{2\sigma^2_{ij}}\right].
\end{equation}
The global minima of such an energy function should be the input 
structure.  

Many have used additional constraining potentials to characterise 
unsampled regions of coordinate space while using molecular dynamics 
\cite{kong96,Eastwood00}.
To characterize the landscape sampled with basin-hopping, we also used a 
structure constraining potential to identify ensembles with fixed but 
varying fractions of native structure.  
Using such a potential allows us to access interesting 
configurations that are unlikely to be thermally sampled. The constraining 
potential also called umbrella 
potentials are centered on different values of an order parameter to 
sample along the collective coordinates. One of the 
collective coordinates is $Q$, an order parameter that 
measures the sequence-dependent structural similarity of two conformations 
by computing the normalized summation of C-alpha pairwise contact differences, 
as defined in Eq.~(\ref{q}):\cite{Eastwood00}
 \begin{equation} 
 \label{q}
 Q=\frac{2}{(N-1)(N-2)}\sum_{i<j-1}
     \exp\left[-\frac{(r_{ij}-r_{ij}^{\rm N})^2}{\sigma_{ij}^2}\right].
\end{equation}
The resulting order parameter ranges from zero, where there is no
similarity between structures, to one, which represents an exact
overlap.
The form of the potential is $E(Q)=2500\epsilon(Q-Q_i)^4$, where 
$Q_i$ may be varied in order to sample different regions of 
the chosen order parameter.  As in equilibrium sampling, 
simulations were initiated at the native 
state and the $Q_i$ parameter was reduced throughout the sampling. 

We have also studied the potential energy landscape when multiple surfaces
are superimposed upon each other by the use of multiple homologous target 
proteins.  
This manipulation of the energy landscape has been shown to further reduce
local energetic frustration that arises from random mutations in the sequence
away from the consensus optimal sequence for a given structure.
By reducing the number of non-native traps, this averaging often 
improves the quality of structure prediction results 
\cite{maxfield79,keasar97,BONN2001c,hardin02}.  
As seen in Eq.~(\ref{seq_ave_eq}), 
the form and the parameters of the energy function are maintained from 
Eqs.~(\ref{amhsrmr}) and (\ref{amhlr}), but the
normalized summation is taken over a set of homologous sequences:
\begin{equation}
 \label{seq_ave_eq}
E_{\rm AM}=-\frac{1}{N_{seq}}\sum^{seq}_{k=1}\sum^{N}_{i<j}E_{int}(P^{k}_{i},P^{k}_{j}).
\end{equation}
Since proteins are not random heteropolymers, the differences in the energy 
function for homologous proteins are randomly distributed, therefore the 
mean over multiple energy functions should have less energetic
variation than the original function.  
Indeed, performing this summation is a way of incorporating to the 
optimisation of the $T_{f}/T_{g}$ criterion into any energy function.
The target sequences of the homologues 
can be identified using PSI-Blast with default parameters \cite{BLAST,bioperl}. Some classes of proteins have a large number
of sequence homologues, therefore performing a multiple sequence alignment
can be impractical.  Removing redundant sequences from within the set of 
identified homologues also 
removes biases that can be introduced where there are few homologues available.
This is done by preventing sequences in the collected sequences from having
greater than 90\% sequence identity.
The remaining sequences are aligned in a multiple sequence alignment
\cite{CLUSTAL}. Gaps within the sequence alignment can be addressed within 
the AMH energy function in a variety of ways. In the present
work, gaps in the target sequence were removed, while gaps within
homologues were completed with residues from the target protein.  While
this procedure may introduce small biases toward the target sequence, it is 
preferable to ignoring the interactions altogether.

Finally, we made several {\it ad hoc}\/ changes to the backbone potential, $E_{\rm back}$. 
Eliminating some compromises necessary for rapid molecular dynamics simulations
allowed the AMH potential to be adapted to basin-hopping.
Another goal was to prevent the over-collapse of the proteins 
by altering the excluded volume energy term, which should reduce the 
number of states available during minimisation. The terms shown in 
Eq.~(\ref{eback}) are used to reproduce the peptide-like conformations in 
the original molecular dynamics energy function:
\begin{equation}
 \label{eback}
E_{\rm back}=E_{\rm ev}+E_{\rm harm}+E_{\rm chain}+E_{\rm chi}+E_{\rm Rama}.
\end{equation}
\(E_{\rm ev}\) maintains a sequence specific excluded
volume constraint between the C$_\alpha$-C$_\alpha$, C$_\beta$-C$_\beta$, O-O, and C$_\alpha$-C$_\beta$
atoms that are separated by less than $r_{\rm ev}$.  Previously
\cite{Eastwood02}, we have seen that modifying $E_{\rm back}$ can produce a
less frustrated energy surface when using thermal equilibrium sampling, 
but slow dynamics was often found to result since
the local barrier heights became too large.  The ability of basin-hopping to
overcome such large, but local barriers allows us therefore to consider a 
potential whose dynamics would otherwise be too slow for molecular 
dynamics.  In the final part of the paper, we altered the excluded 
volume term, as shown in Eq.~(\ref{ev_eq}) to prevent over-collapse:
\begin{equation}
 \label{ev_eq}
E_{\rm ev}=\epsilon\lambda_{\rm RV}^{\rm C}\sum_{x,y}\sum_{i<j}\theta(r_{\rm ev}^{\rm C}(j-i)-r_{{\rm C}^x_i{\rm C}^y_j})(r_{\rm ev}^{\rm C}(j-i)-r_{{\rm C}^x_i{\rm C}^y_j})^2 +\epsilon\lambda_{\rm ev}^{\rm O}\sum_{i<j}\theta(r_{\rm ev}^{\rm O}-r_{{\rm O}_i{\rm O}_j})(r_{\rm ev}^{\rm O}-r_{{\rm O}_i{\rm O}_j})^2,
\end{equation}
by changing the default molecular dynamics parameters, 
$\lambda_{\rm EV}^{\rm C}=20$, $\lambda_{\rm EV}^{\rm O}=20$,
$r_{\rm ev}^{\rm C}(j-i<5)=3.85\,$\AA, $r_{\rm ev}^{\rm C}(j-i\ge 5)=4.5\,$\AA,
and $r_{\rm ev}^{\rm O}=3.5\,$\AA, to 
$\lambda_{\rm EV}^{\rm C}=250$, 
$\lambda_{\rm EV}^{\rm O}=250$, $r_{\rm ev}^{\rm C}(j-i<5)=3.85\,$\AA, 
$r_{\rm ev}^{\rm C}(j-i\ge 5)=3.85\,$\AA, and $r_{\rm ev}^{\rm O}=3.85\,$\AA.
The force constant are over an order of magnitude
larger than those used in the molecular dynamics, and the radii
of the C$_\alpha$, C$_\beta$, and O atoms are also 10\%\ larger than 
previously used values. This increase in excluded volume slows the onset of
chain collapse, but improves steric interactions. The other change to 
the backbone potential involves terms which maintain chain connectivity.
In molecular dynamics annealing, covalent bonds are preserved 
using the SHAKE algorithm \cite{Ryckaert77}, which permits an increase of the 
molecular dynamics time step.  For basin-hopping in all parts of this paper, we removed the SHAKE method and replaced it with a 
harmonic potential, $E_{\rm harm}$, between the C$_{\alpha}$-C$_{\alpha}$, 
C$_{\alpha}$-C$_{\beta}$, and C$_{\alpha}$-O atoms.  This replacement permits 
the location of local minima without requiring an internal coordinate 
transformation, and avoids discontinuous gradients.
When minimised, the additional harmonic terms typically contribute only 
only .015 \,$k_BT$ per bond.  The remaining terms of the original 
backbone potential are maintained.  Depending on the sidechain, 
the neighbouring residues in sequence sterically limit the variety 
of positions the backbone atoms can occupy, as evidenced in a Ramachandran 
plot \cite{Rama}. This
distribution of coordinates is reinforced by a potential, \(E_{\rm Rama}\),
with artificially low barriers to encourage rapid local movements.
The planarity of the peptide bond is ensured by a harmonic potential, 
\(E_{\rm chain}\).  The chirality of the C$_\alpha$ centres is maintained 
using the scalar triple product of neighbouring unit vectors of carbon and nitrogen 
bonds, \(E_{\rm chi}\).

For basin-hopping simulations, whose algorithm is outlined in
Fig.~\ref{pseudocode} the most important sampling parameters 
are the temperature used in the accept/reject steps for local minima $T_{bh}$,
and the maximum step size for perturbations of the Cartesian 
coordinates $d$. A higher temperature allows transitions to an increased 
energy minima to be accepted, and also creates a larger the number
of iterations typically required to minimise the greater perturbed 
configurations. Too high a temperature leads to insufficient exploration of 
low-energy regions.  The temperature ($T_{bh}$) for these simulations was 10 $T_{amc}$. 
Lower temperatures resulted in slower escape rates from low energy traps, while
higher temperatures prevented adequate exploration of low energy regions.
The step size needs to be large enough to move the configuration into 
the basin of attraction of one local minimum to a neighbouring one, but not be 
so large that the new minimum is unrelated to the previous state.  
Every Cartesian coordinate was displaced up to a maximum step size($d$) of 
$0.75\,$\AA, the optimum value determined from preliminary tests.
Each run consisted of 2500 basin-hopping steps saving structures every 5
basin-hopping steps.  The convergence condition ($\delta E_{min}$) on the root-mean-square
(RMS) gradient for each minimisation was set to $10^{-3}\,\epsilon$, and
the 5 lowest-lying minima from each run were subsequently converged
more tightly ($\delta E_{final}$)  to an RMS gradient of $10^{-5}\epsilon$. 
It is important to note that basin-hopping does not provide equilibrium thermodynamic sampling.
In structure prediction there, however, is no rigorous need
for the search to obey detailed balance since the global energy minimum is the 
primary interest.  Basin-hopping provides a means for the
optimal global search of the energy
landscape, however other methods must be used when calculating free energy and
entropy.

In previous structure prediction studies with the AMH, low energy
structures were identified using off-lattice Langevin dynamics with
simulated annealing, employing a linear annealing schedule of 10000
steps from a temperature of 2.0 to 0.0, starting from a random configuration
\cite{FriedrichsMS90}.  The number and length of simulations needed in 
both strategies were determined by the number of uncorrelated structures
encountered.  The current basin-hopping method with the AMC energy function
encounters roughly one deep trap per run.  In order to sample 100 independent
structures in molecular dynamics 20 separate runs were needed, because
simulated annealing samples about five independent states before the glass
transition temperature is encountered, as measured by the rapid
decay of structural correlations. We compared several $\alpha$ helical proteins,
both from within and outside the training set of the AMH energy function.

\section{Results and Discussion}
\begin{figure}[htb]
  \centering
\psfrag{Q}[tc][bc]{$Q$}
\psfrag{E(Q)}[cr][cr]{$E(Q)/\epsilon$}
\psfrag{0}[tc][tc]{$0$}
\psfrag{0.2}[tc][tc]{$0.2$}
\psfrag{0.4}[tc][tc]{$0.4$}
\psfrag{0.6}[tc][tc]{$0.6$}
\psfrag{0.8}[tc][tc]{$0.8$}
\psfrag{1}[tc][tc]{$1$}
\psfrag{-400}[cr][cr]{$-400$}
\psfrag{-350}[cr][cr]{$-350$}
\psfrag{-300}[cr][cr]{$-300$}
\psfrag{-250}[cr][cr]{$-250$}
\psfrag{-200}[cr][cr]{$-200$}
\psfrag{-150}[cr][cr]{$-150$}
\psfrag{-100}[cr][cr]{$-100$}
  \includegraphics[width=0.6\textwidth]{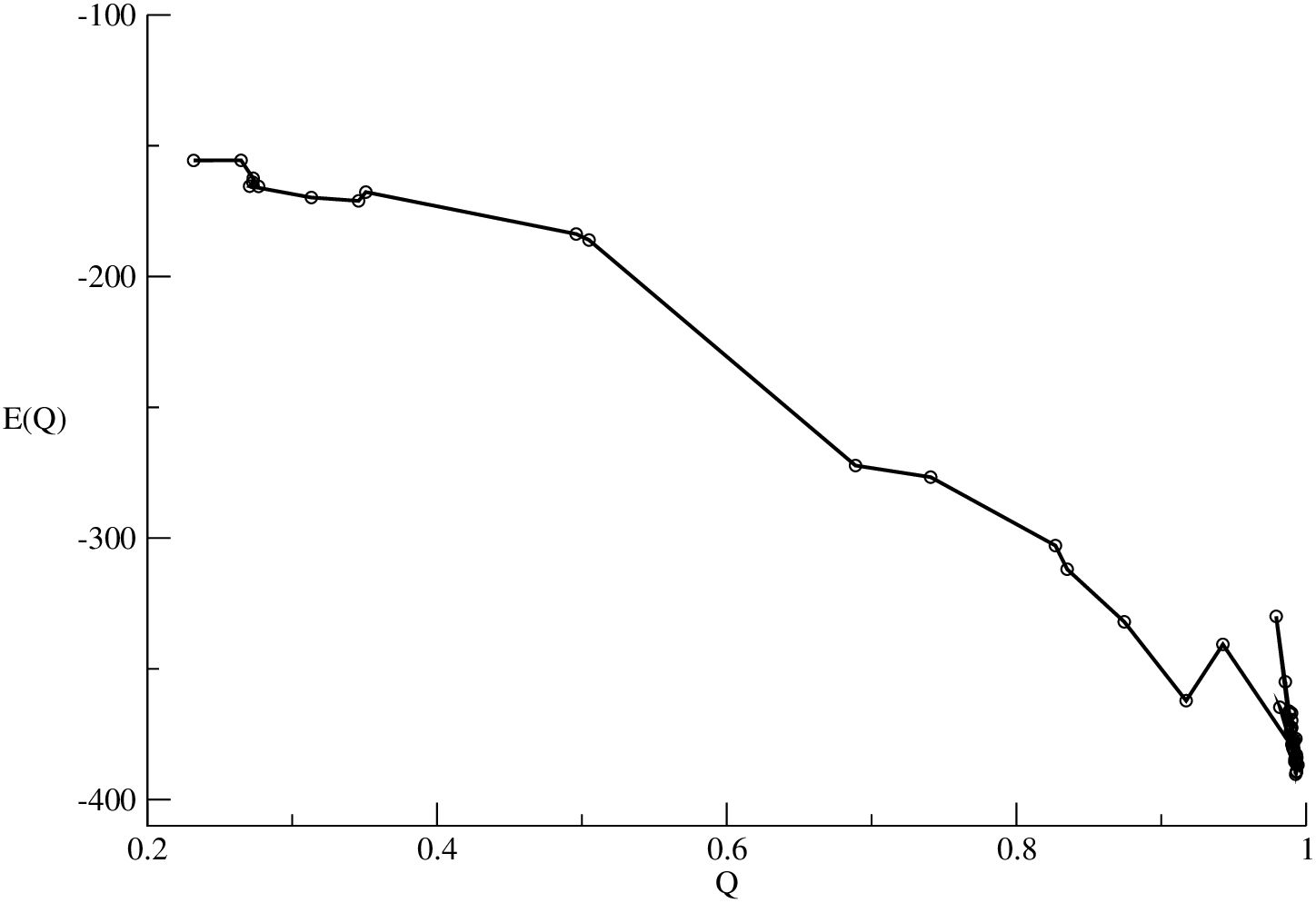}
  \caption{\label{Go_q}
}
\end{figure}

We performed initial calculations with a 
G\=o potential for 
the 434 repressor (protein data bank (PDB \cite{PDB}) ID 1r69).
In Fig.~\ref{Go_q} we show this model accurately represents the native
basin.
Steps where the energy increases are allowed by the sampling method 
and are not examples of frustration.
Studies on the G\=o model provide a useful benchmark for 
comparing the computer time required for the different global optimisation 
strategies.  Using the sampling parameters used in this report, we compared 
the time for initial collapse between the molecular dynamics and 
basin-hopping runs.  The initial collapse required about 7 minutes for 
the annealing runs and 31 minutes for basin-hopping on a desktop computer.
However, these values do not reflect the actual performance of the two 
approaches in locating global minima, which will depend upon the move sets,
step size, temperature, and convergence criteria. 

While using the AMC structure prediction Hamiltonian, we found that 
basin-hopping was often able to locate lower energy structures and also 
identified minima that have greater structural overlap with the 
native state than annealing. These results are produced for structure 
predictions for proteins both inside and outside the training set, as 
demonstrated in Table~\ref{gmin_compare}.  The first three proteins 
(PDB ID 1r69, 3icb, 256b) in Table~\ref{gmin_compare} 
are in the training set of the Hamiltonian {\cite{Hardin00}}, while the other 
three are not, and can therefore be considered as predictions from 
the algorithm.  The minima located with basin-hopping show an increase 
in structural overlap with the native state [Eq.~(\ref{q})] 
when compared to the Langevin dynamics approach. $Q$ scores of 0.4 for 
single domain proteins generally correspond to a low resolution root mean 
square deviation (RMSD) of around 5 \AA  or better. $Q$ scores of 0.5 and 
higher have still more accurate tertiary packing and are of comparable 
quality to the experimentally derived models.  The high quality 
structures obtained suggest the form of the backbone terms is 
appropriate, since the physically correct stereochemistry is reproduced. 
Lower energy structures are sampled by basin-hopping for the non-training 
set proteins, but the structural overlap improvement found in these deeper 
minima was smaller.  Larger proteins pose a greater challenge for 
basin-hopping with this Hamiltonian due to the random steps in 
Cartesian coordinates.  Dihedral coordinate move steps would probably be 
more efficient, and will be considered in future work.

\begin{figure}[htb]
\psfrag{Q}[tc][bc]{$Q$}
\psfrag{E(Q)}[cr][cr]{$E(Q)/\epsilon$}
\psfrag{0.1}[tc][tc]{$0.1$}
\psfrag{0.2}[tc][tc]{$0.2$}
\psfrag{0.3}[tc][tc]{$0.3$}
\psfrag{0.4}[tc][tc]{$0.4$}
\psfrag{0.5}[tc][tc]{$0.5$}
\psfrag{0.6}[tc][tc]{$0.6$}
\psfrag{-500}[cr][cr]{$-500$}
\psfrag{-400}[cr][cr]{$-400$}
\psfrag{-300}[cr][cr]{$-300$}
\psfrag{-200}[cr][cr]{$-200$}
\psfrag{-100}[cr][cr]{$-100$}
\psfrag{0}[cr][cr]{$0$}
\psfrag{100}[cr][cr]{$100$}
\psfrag{Structures Encountered During Basin Hopping}[cr][cr]{$Structures Encountered During Basin Hopping$}
  \centerline{\includegraphics[width=0.6\textwidth]{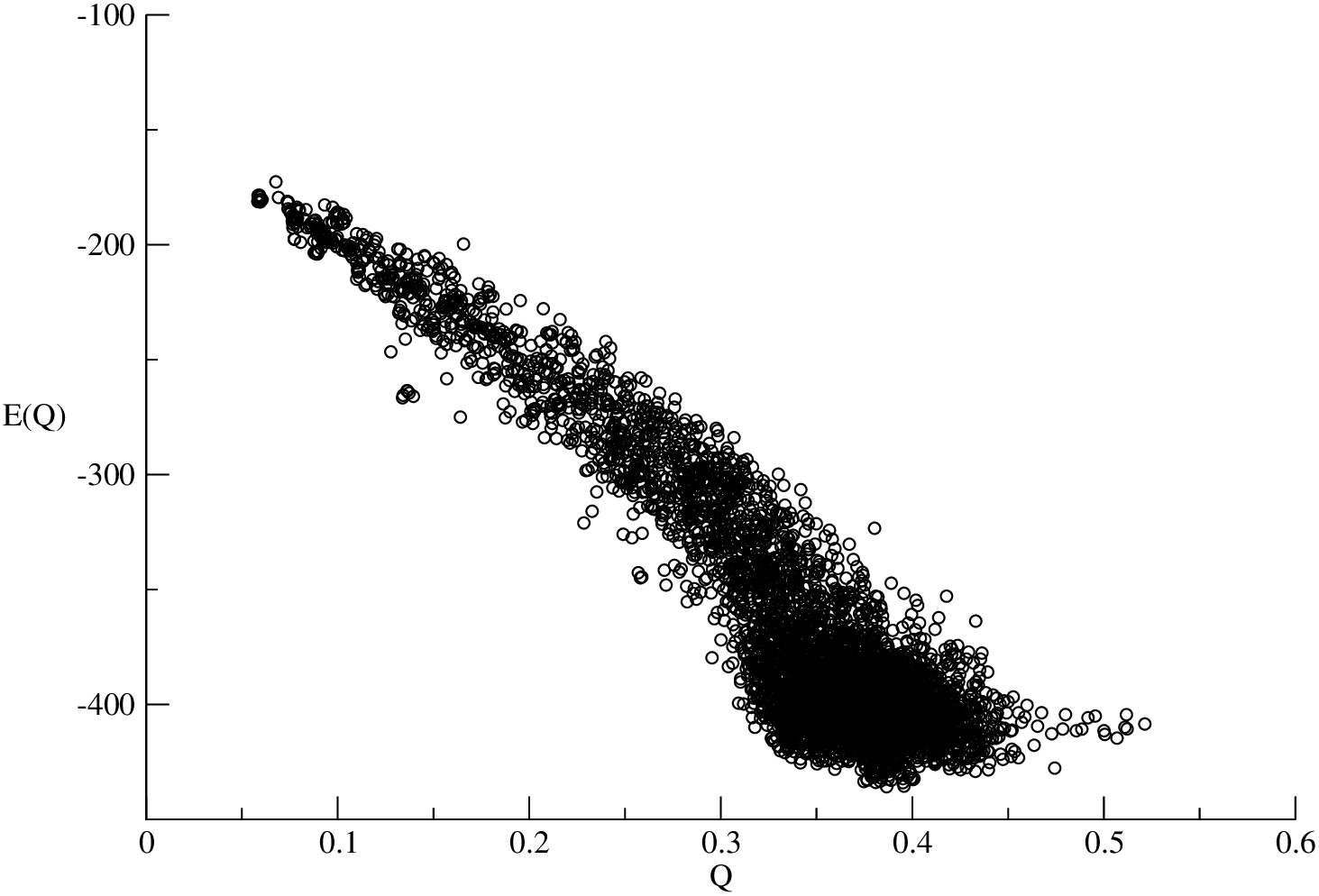}}
  \vskip 2.0 cm
\psfrag{Q}[tc][bc]{$Q$}
\psfrag{E(Q)}[cr][cr]{$E(Q)/\epsilon$}
\psfrag{0}[tc][tc]{$0.0$}
\psfrag{0.1}[tc][tc]{$0.1$}
\psfrag{0.2}[tc][tc]{$0.2$}
\psfrag{0.3}[tc][tc]{$0.3$}
\psfrag{0.4}[tc][tc]{$0.4$}
\psfrag{0.5}[tc][tc]{$0.5$}
\psfrag{0.6}[tc][tc]{$0.6$}
\psfrag{-500}[cr][cr]{$-500$}
\psfrag{-400}[cr][cr]{$-400$}
\psfrag{-300}[cr][cr]{$-300$}
\psfrag{-200}[cr][cr]{$-200$}
\psfrag{-100}[cr][cr]{$-100$}
\psfrag{0}[cr][cr]{$0$}
\psfrag{100}[cr][cr]{$100$}
\psfrag{Structures Encountered During Simulated Annealing}[cr][cr]{$Structures Encountered During Simulated Annealing$}
  \centerline{\includegraphics[width=0.6\textwidth]{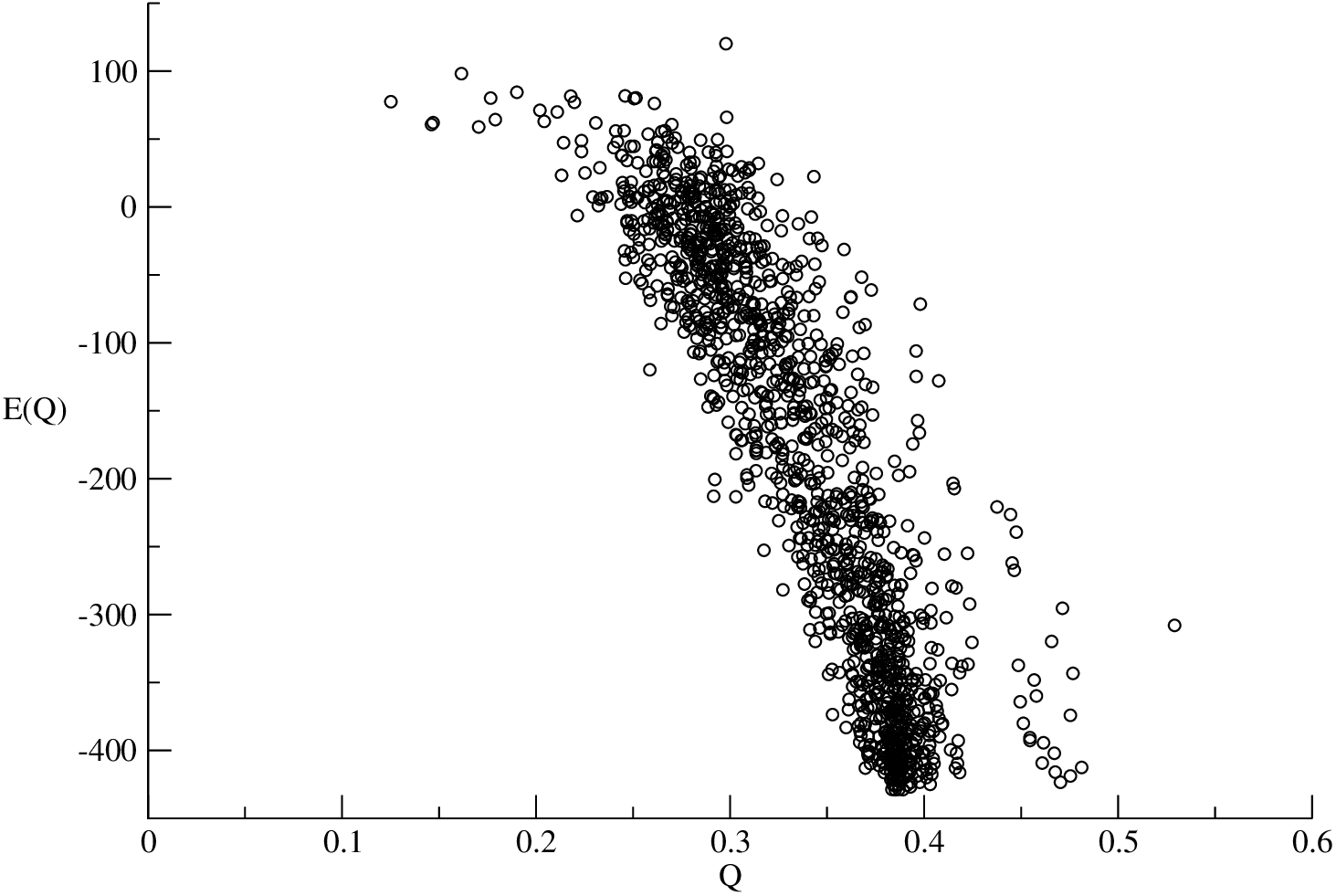}}
 \caption{\label{gmin_anneal}
}
\end{figure}

\begin{figure}[htp]
\psfrag{Q}[tc][bc]{$Q$}
\psfrag{E(Q)}[cr][cr]{$E(Q)/\epsilon$}
\psfrag{Index}[cr][cr]{Index}
\psfrag{0}[tc][tc]{$0.0$}
\psfrag{0.2}[tc][tc]{$0.20$}
\psfrag{0.25}[tc][tc]{$0.25$}
\psfrag{0.3}[tc][tc]{$0.30$}
\psfrag{0.35}[tc][tc]{$0.35$}
\psfrag{0.4}[tc][tc]{$0.40$}
\psfrag{0.45}[tc][tc]{$0.45$}
\psfrag{0.5}[tc][tc]{$0.50$}
\psfrag{0.55}[tc][tc]{$0.55$}
\psfrag{0.6}[tc][tc]{$0.60$}
\psfrag{0}[tc][tc]{$0$}
\psfrag{20}[tc][tc]{$20$}
\psfrag{40}[tc][tc]{$40$}
\psfrag{60}[tc][tc]{$60$}
\psfrag{80}[tc][tc]{$80$}
\psfrag{100}[tc][tc]{$100$}
\psfrag{-420}[cr][cr]{$-420$}
\psfrag{-400}[cr][cr]{$-400$}
\psfrag{-380}[cr][cr]{$-380$}
\psfrag{-360}[cr][cr]{$-360$}
\psfrag{-340}[cr][cr]{$-340$}
\psfrag{-320}[cr][cr]{$-320$}
  \centerline{\includegraphics[width=0.6\textwidth]{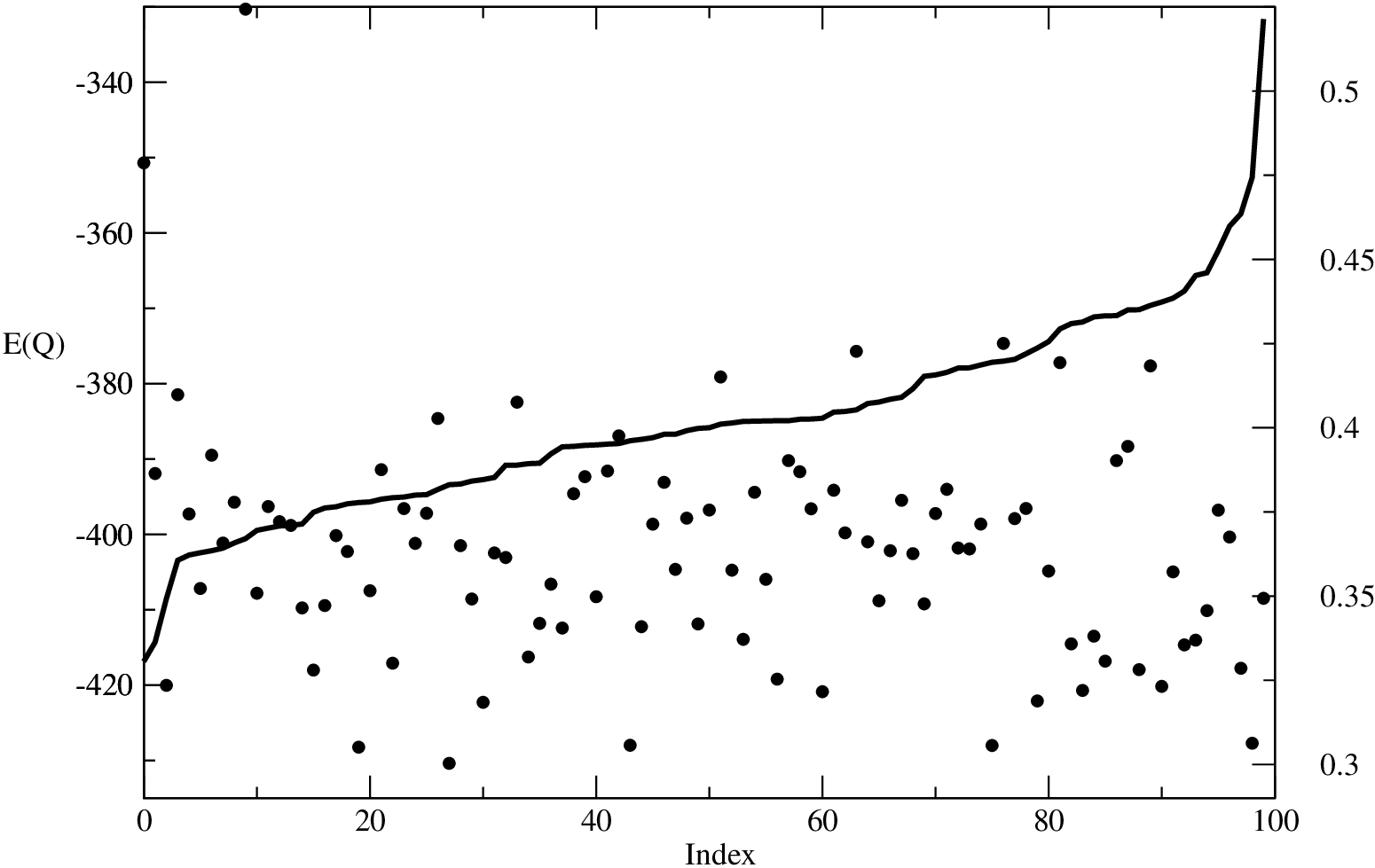}}
  \vskip 2.0 cm
\psfrag{Q}[tc][bc]{$Q$}
\psfrag{E(Q)}[cr][cr]{$E(Q)/\epsilon$}
\psfrag{Index}[cr][cr]{Index}
\psfrag{0}[tc][tc]{$0.0$}
\psfrag{0.1}[tc][tc]{$0.10$}
\psfrag{0.15}[tc][tc]{$0.15$}
\psfrag{0.2}[tc][tc]{$0.20$}
\psfrag{0.25}[tc][tc]{$0.25$}
\psfrag{0.3}[tc][tc]{$0.30$}
\psfrag{0.35}[tc][tc]{$0.35$}
\psfrag{0.4}[tc][tc]{$0.40$}
\psfrag{0.45}[tc][tc]{$0.45$}
\psfrag{0.5}[tc][tc]{$0.50$}
\psfrag{0}[tc][tc]{$0$}
\psfrag{20}[tc][tc]{$20$}
\psfrag{40}[tc][tc]{$40$}
\psfrag{60}[tc][tc]{$60$}
\psfrag{80}[tc][tc]{$80$}
\psfrag{100}[tc][tc]{$100$}
\psfrag{-400}[cr][cr]{$-400$}
\psfrag{-450}[cr][cr]{$-450$}
\psfrag{-300}[cr][cr]{$-300$}
\psfrag{-350}[cr][cr]{$-350$}
  \centerline{\includegraphics[width=0.6\textwidth]{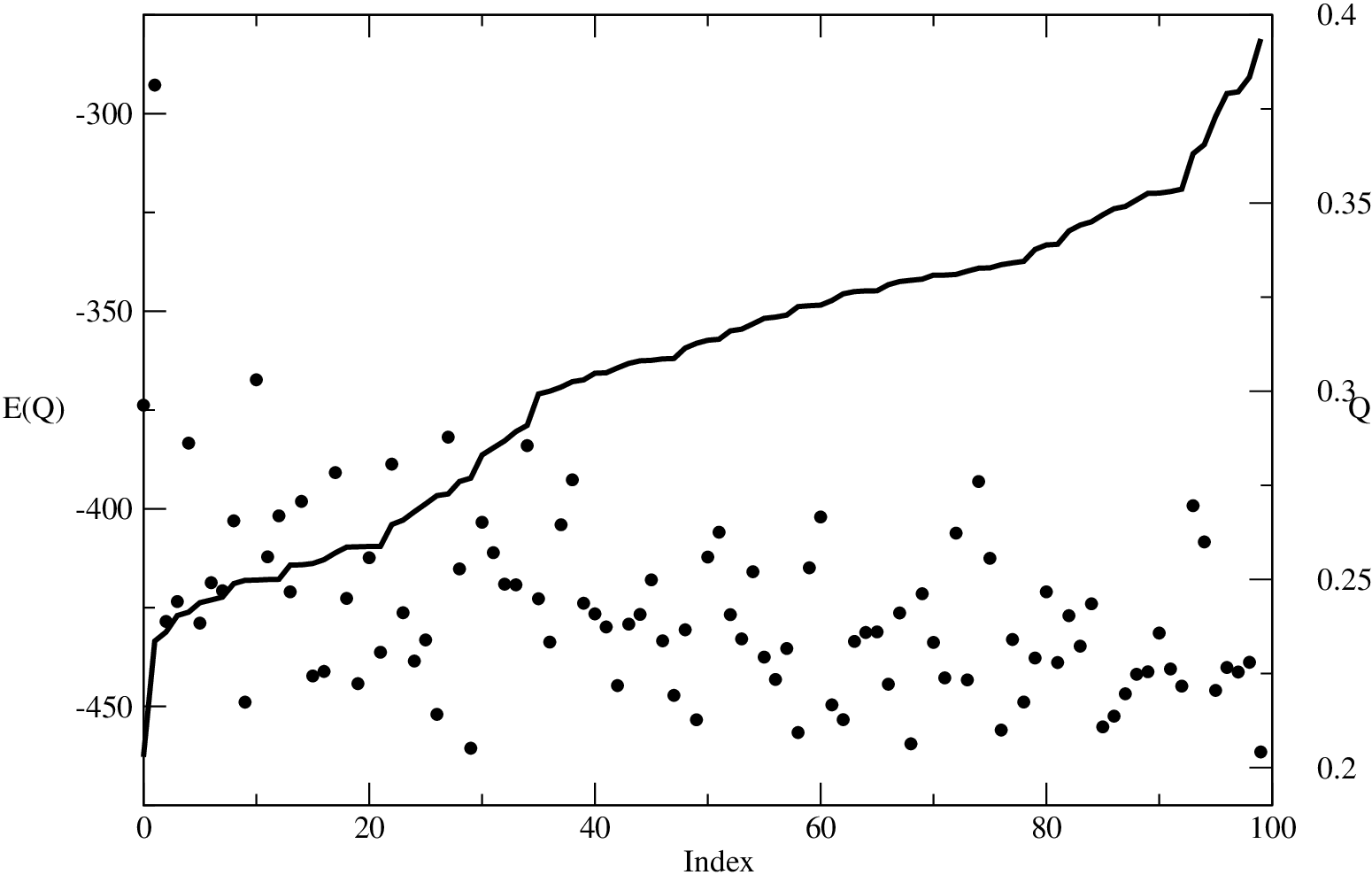}}
  \caption{\label{yield_plot}
}
\end{figure}

The distribution of minima encountered from multiple simulations for both 
search methods is shown in Fig.~\ref{gmin_anneal} where a 
greater density of high quality structures is obtained by the basin-hopping
algorithm.  The potential energy surface still includes, therefore, 
significant residual frustration in the near-native 
basin in the form of low-lying minima separated by relatively high 
barriers. Without the parameter optimisation to reduce frustration, 
folding would exhibit more pronounced glassy characteristics.
Most of the cooperative folding occurs during collapse until $Q$ values of 
around 0.4 are reached.  While the structures from simulated annealing 
are accurate enough for functional determination,
we see basin-hopping can better overcome barriers that are created 
after collapse. The density of the high quality structures is also 
important for 
post-simulation k-means clustering analysis {\cite{Baker_1998}}.  Another 
way of representing the data of a set of independent basin-hopping 
simulations is by selecting the lowest energy structures from each 
simulation of the 434 repressor (PDB ID 1r69) and HDEA (PDB ID 1bg8) 
proteins and ordering them with respect to their structural overlap.  
As shown in 
Fig.~\ref{yield_plot}, the protein in the training set (434 repressor) 
produces better results than the non-training protein, as expected.  

We have decomposed the different energy terms in the Hamiltonian in 
Table~\ref{terms}, to examine which interactions are most effectively 
minimised.  The AMC potential has three different distance 
classes in terms of sequence separation, 
and these are defined as short ($\left\vert i-j\right\vert<5$), 
medium ($5\le\left\vert i-j\right\vert\le12$), and long 
($\left\vert i-j\right\vert>12$).  Most importantly, the long-range AMH 
interactions are 
successfully minimised in the basin-hopping runs, due to the ability of 
basin-hopping to overcome large energetic barriers.  This term will govern the 
quality of structures sampled using an approximately smooth energy landscape. 
The other terms that define secondary structure formation are not 
as well minimised. This result is due to the disruption of helices by the 
random Cartesian perturbation move steps. These move steps benefit favorable 
steric packing and therefore do well at minimising the excluded volume 
energy term of the Hamiltonian. A combined 
minimisation approach might be more efficient, where larger dihedral steps 
could be made early during minimisation to sample a wider number of structures,
followed by random Cartesian steps to optimise the steric interactions.

\begin{figure}[htb]
\psfrag{Q}[tc][bc]{$Q$}
\psfrag{E(Q)}[cr][cur]{$E(Q)/\epsilon$}
\psfrag{0}[tc][tc]{$0$}
\psfrag{0.1}[tc][tc]{$0.1$}
\psfrag{0.2}[tc][tc]{$0.2$}
\psfrag{0.3}[tc][tc]{$0.3$}
\psfrag{0.4}[tc][tc]{$0.4$}
\psfrag{0.5}[tc][tc]{$0.5$}
\psfrag{0.6}[tc][tc]{$0.6$}
\psfrag{0.7}[tc][tc]{$0.7$}
\psfrag{0.8}[tc][tc]{$0.8$}
\psfrag{0.9}[tc][tc]{$0.9$}
\psfrag{1}[tc][tc]{$1$}
\psfrag{-440}[cr][cr]{$-440$}
\psfrag{-430}[cr][cr]{$-430$}
\psfrag{-420}[cr][cr]{$-420$}
\psfrag{-410}[cr][cr]{$-410$}
\psfrag{-400}[cr][cr]{$-400$}
\psfrag{-390}[cr][cr]{$-390$}
\psfrag{-380}[cr][cr]{$-380$}
\psfrag{-360}[cr][cr]{$-360$}
\psfrag{-340}[cr][cr]{$-340$}
  \centerline{\includegraphics[width=0.6\textwidth]{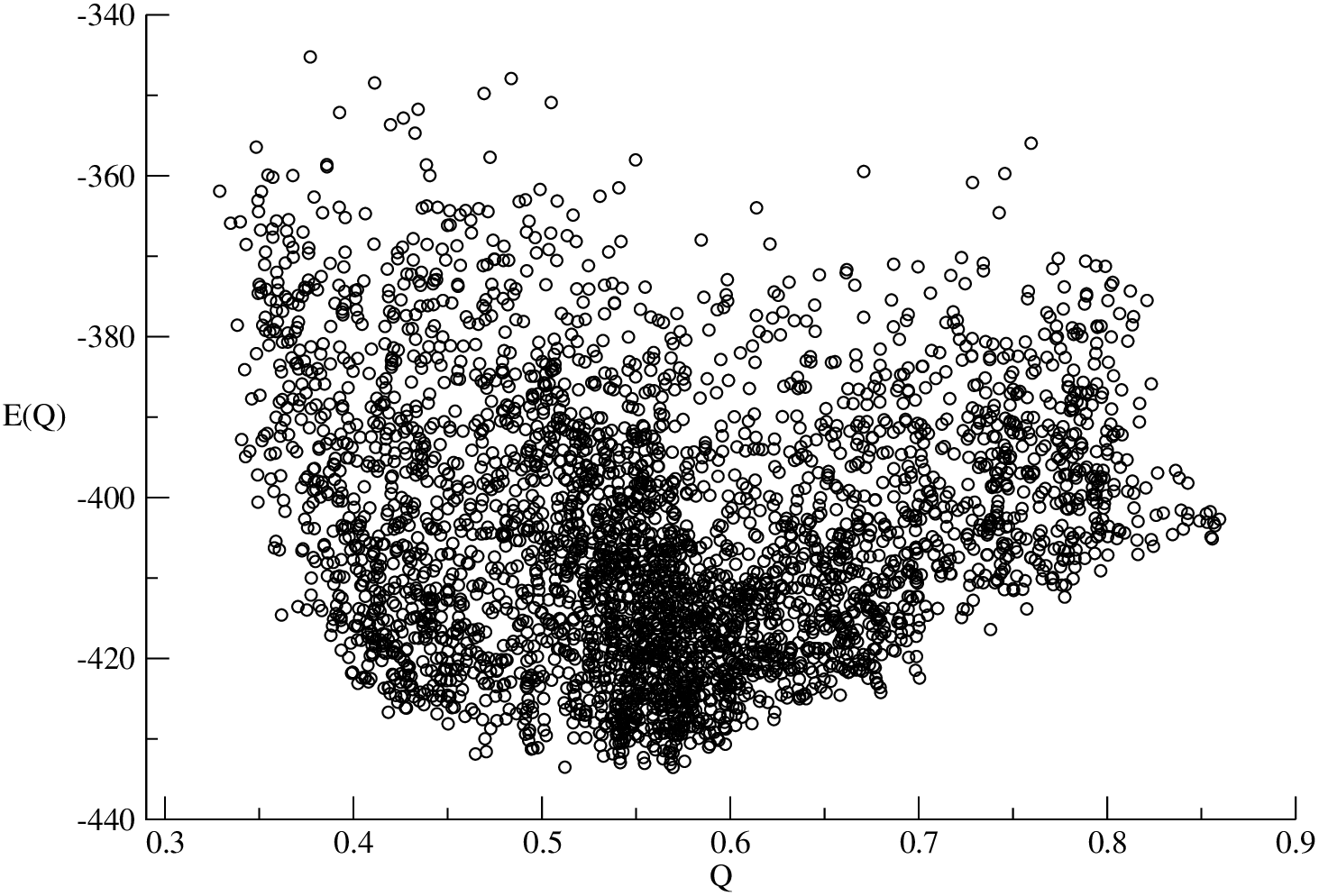}}
  \vskip 2.0 cm
\psfrag{-760}[cr][cr]{$-760$}
\psfrag{-750}[cr][cr]{$-750$}
\psfrag{-740}[cr][cr]{$-740$}
\psfrag{-730}[cr][cr]{$-730$}
\psfrag{-720}[cr][cr]{$-720$}
\psfrag{-710}[cr][cr]{$-710$}
\psfrag{-700}[cr][cr]{$-700$}
\psfrag{-680}[cr][cr]{$-680$}
\psfrag{-660}[cr][cr]{$-660$}
  \centerline{\includegraphics[width=0.6\textwidth]{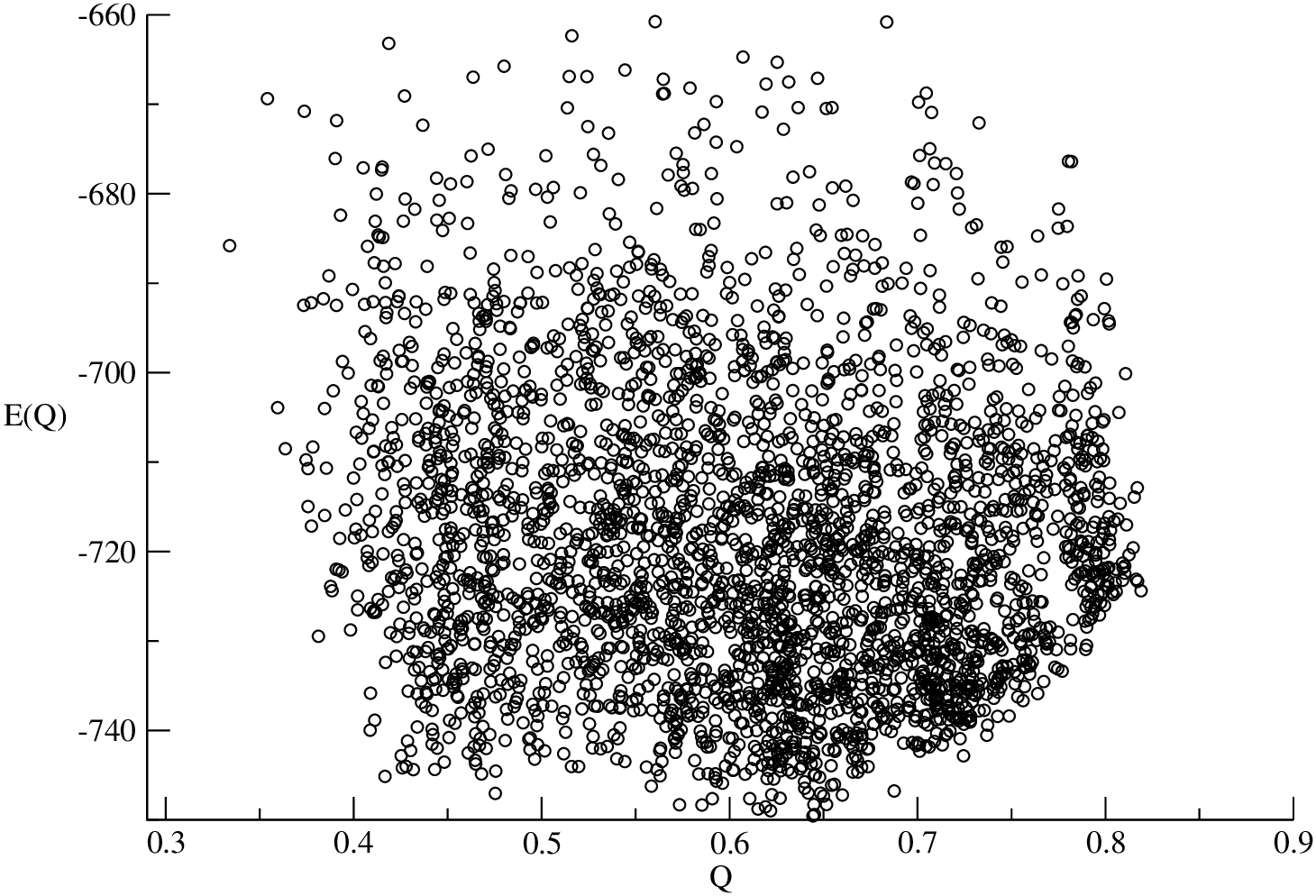}}
  \caption{\label{q_const_256b}
}
\end{figure}

While we sampled high quality structures, we would like to confirm that we
have completely sampled the global minima of the energy surface.
To access these unsampled states we used
umbrella potentials.  When constraining a set of simulations to
different values of $Q$, we have obtained energy minima
for cytochrome c, roughly $15\,\epsilon$ deeper than those from
unconstrained minimisations starting
with a randomized structure, as shown in Fig.~\ref{q_const_256b}. For the
434 repressor the minima obtained from randomized states and those found
with the $Q$ constraints applied differ by only a few $k_BT$.
This shows basin-hopping does indeed act as a global optimisation method,
by accurately identifying the global energy minimum from multiple
independent unconstrained simulation.
This behavior is predictable from the choices that governed
the design of the Hamiltonian.  Low energy barriers
between structures are desirable during a molecular dynamics simulation because
they accelerate the dynamics.  However, for basin-hopping these low barriers
encourage tertiary contact formation before secondary structure units condense
for sequences greater than 110 amino acids.
 
\subsection*{Superposition of Multiple Energy Landscapes}

\begin{figure}[htp]
\psfrag{Q}[tc][bc]{$Q$}
\psfrag{E(Q)}[cr][cr]{$E(Q)/\epsilon$}
\psfrag{0}[tc][tc]{$0.0$}
\psfrag{0.1}[tc][tc]{$0.1$}
\psfrag{0.2}[tc][tc]{$0.2$}
\psfrag{0.3}[tc][tc]{$0.3$}
\psfrag{0.4}[tc][tc]{$0.4$}
\psfrag{0.5}[tc][tc]{$0.5$}
\psfrag{0.6}[tc][tc]{$0.6$}
\psfrag{-100}[cr][cr]{$-100$}
\psfrag{-200}[cr][cr]{$-200$}
\psfrag{-300}[cr][cr]{$-300$}
\psfrag{-400}[cr][cr]{$-400$}
\psfrag{-500}[cr][cr]{$-500$}
  \centerline{\includegraphics[width=0.6\textwidth]{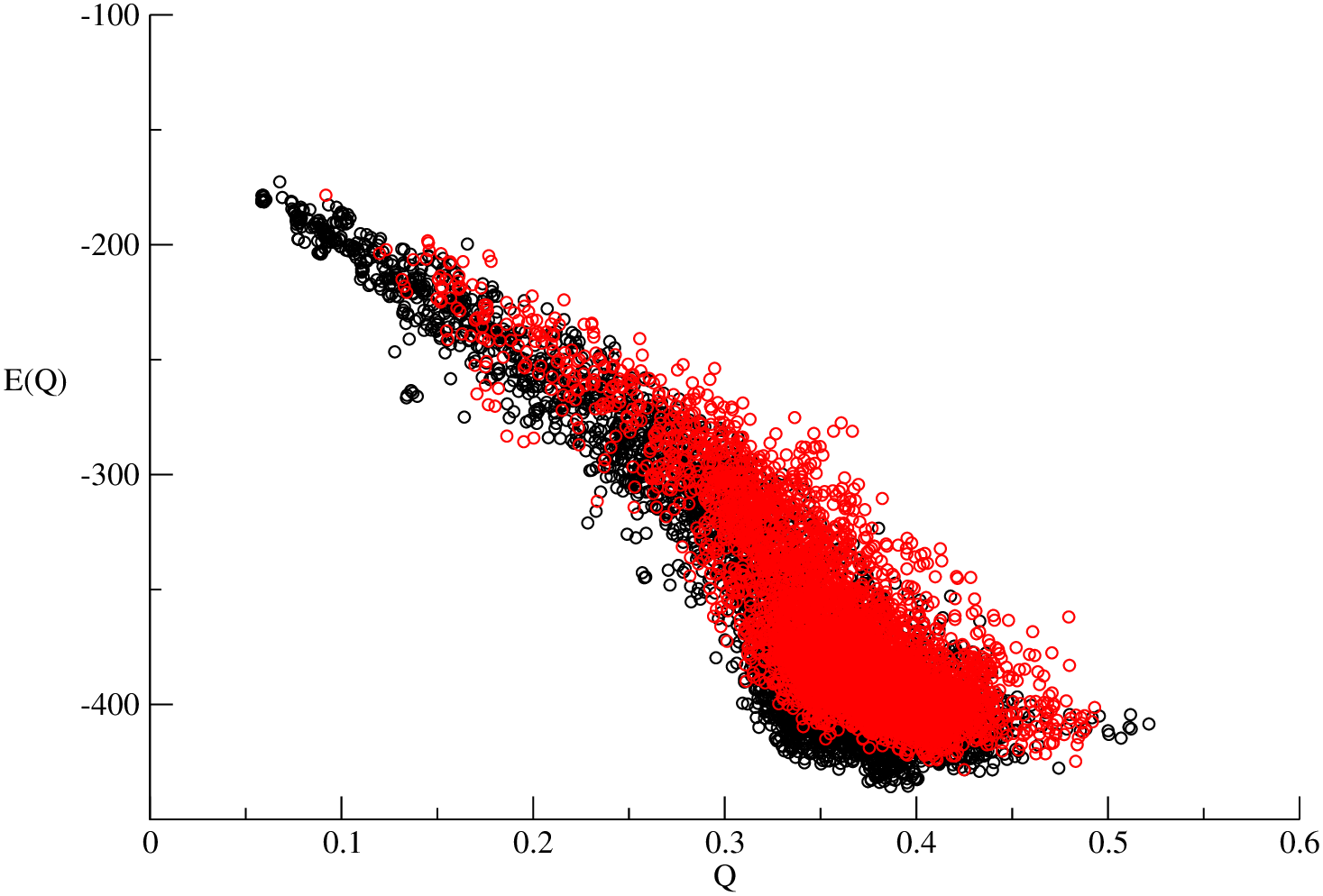}}
\vskip 2.0 cm
\psfrag{Q}[tc][bc]{$Q$}
\psfrag{E(Q)}[cr][cr]{$E(Q)/\epsilon$}
\psfrag{0}[tc][tc]{$0.0$}
\psfrag{0.1}[tc][tc]{$0.1$}
\psfrag{0.2}[tc][tc]{$0.2$}
\psfrag{0.3}[tc][tc]{$0.3$}
\psfrag{0.4}[tc][tc]{$0.4$}
\psfrag{0.5}[tc][tc]{$0.5$}
\psfrag{-100}[cr][cr]{$-100$}
\psfrag{-200}[cr][cr]{$-200$}
\psfrag{-300}[cr][cr]{$-300$}
\psfrag{-400}[cr][cr]{$-400$}
\psfrag{-500}[cr][cr]{$-500$}
  \centerline{\includegraphics[width=0.6\textwidth]{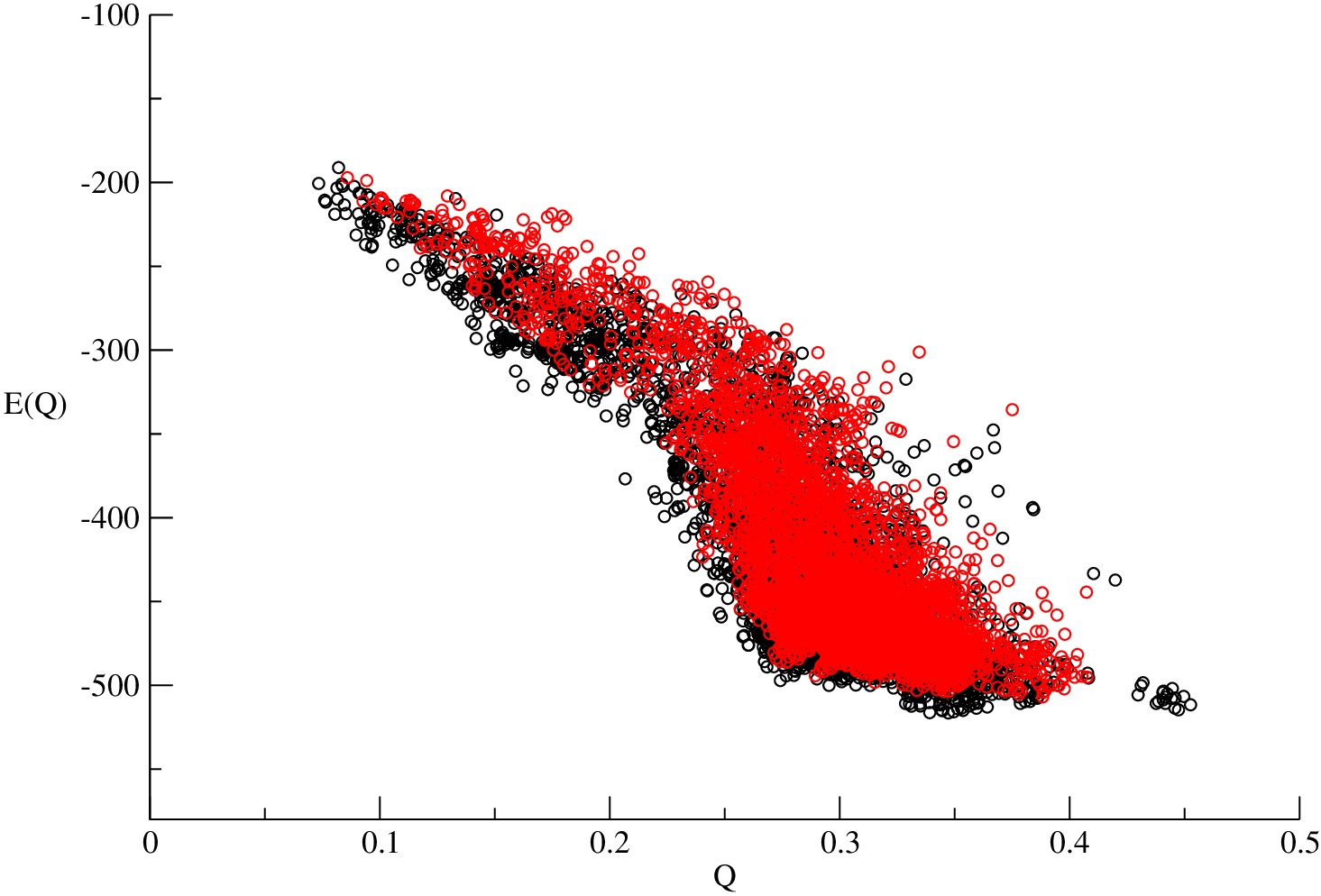}}
  \caption{\label{seq_ave} 
}
\end{figure}

Constructing a Hamiltonian by calculating the arithmetic average of the 
potential over a set of homologous sequences increased the quality of  
predictions in both equilibrium and annealing simulations.  
We have found this approach also improved the performance in 
basin-hopping simulations.  For two different proteins, 100 independent 
basin-hopping runs were performed with both the standard and 
sequence-averaged Hamiltonians.  By the superposition of multiple energy 
landscapes we saw a reduction in the number of competing low energy traps 
around $Q$ values of $0.3$ for both the 434 repressor and uteroglobin 
(PDB ID 1UTG), as shown in Fig.~\ref{seq_ave}.  Improvement of structure 
prediction Hamiltonians can be statistically 
described by the average energy gap between the native basin and a set of 
unfolded structures, and by the roughness of the energy surface, which 
corresponds to the variance of the energy.  
The sequence based energy function summations limited the 
energetic variance of the sampled landscapes, thereby reducing the 
glass transition temperature.  This improvement, even at the low temperatures  
sampled in basin-hopping, is predicted from theory, but difficult to 
observe in conventional equilibrium simulations due to the emergent glassy 
dynamics, which slows the kinetics.  The energy gap improvement was smaller 
than the reduction of the energetic variation of the Hamiltonian. 
In terms of the goal of maximizing the ratio of $T_{f}/T_{g}$, this increase
came primarily from to reducing the glass transition temperature $T_{g}$. 
In the low energy region we saw fewer competing states, and an increased 
correlation between ${E}$ and ${Q}$ for the sequence-averaged Hamiltonian 
compared to the original Hamiltonian.  For the 434 
repressor the lowest energy structure had the highest $Q$ value encountered.

\subsection*{Characterisation of Polymer Collapse}

\begin{figure}[htb]
\psfrag{Q}[tc][bc]{$Q$}
\psfrag{Monte Carlo Steps}[tc][bc]{Basin-Hopping Steps}
\psfrag{Ratio Model to Native Contacts}[cr][cr]{Total Contacts/Native Contacts}
\psfrag{0}[tc][tc]{$0$}
\psfrag{0.6}[cr][cr]{$0.6$}
\psfrag{0.8}[cr][cr]{$0.8$}
\psfrag{1}[cr][cr]{$1.0$}
\psfrag{1.2}[cr][cr]{$1.2$}
\psfrag{1.4}[cr][cr]{$1.4$}
\psfrag{10}[tc][tc]{$10$}
\psfrag{20}[tc][tc]{$20$}
\psfrag{30}[tc][tc]{$30$}
\psfrag{40}[tc][tc]{$40$}
\psfrag{50}[tc][tc]{$50$}
  \centering
  \includegraphics[width=0.6\textwidth]{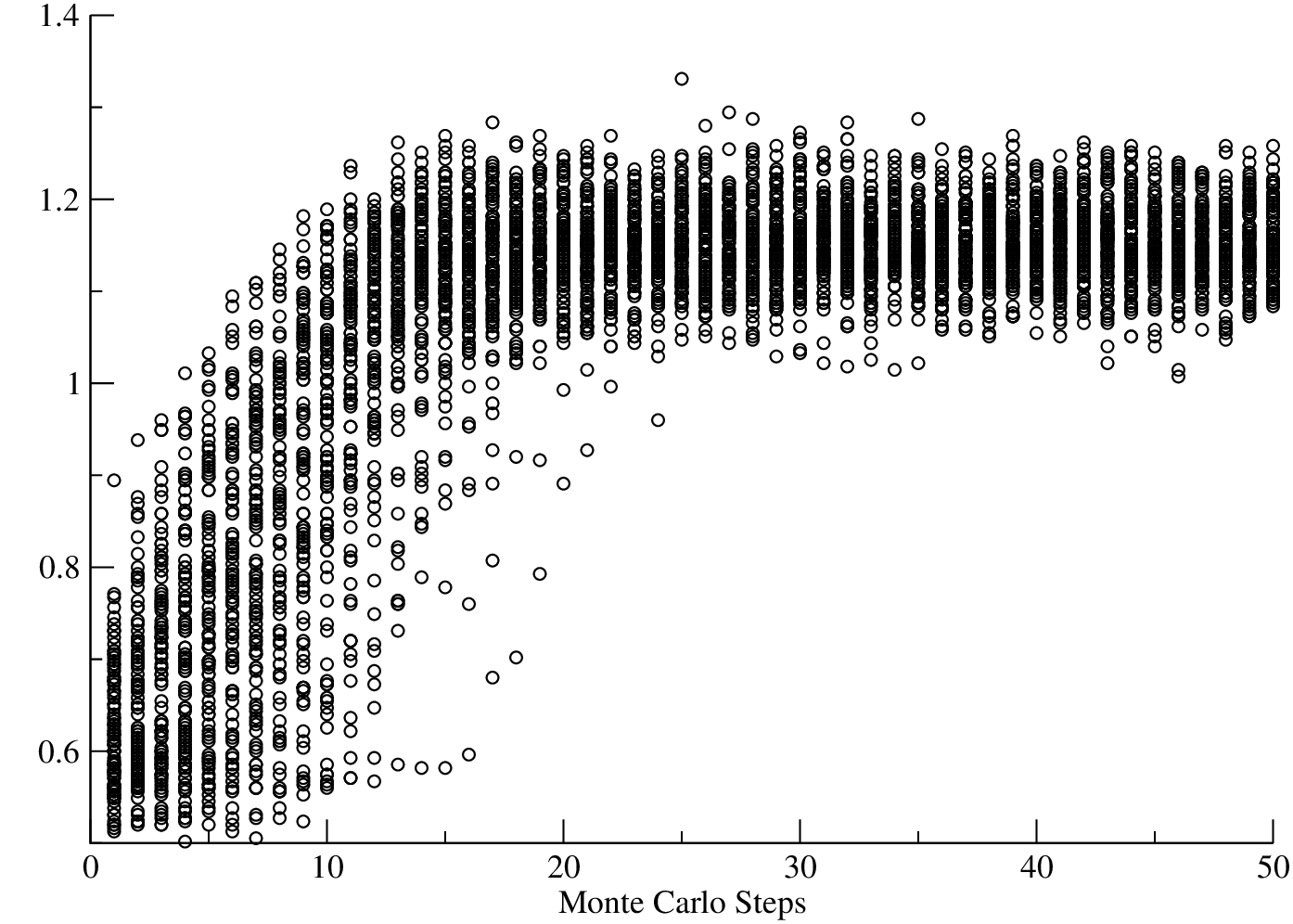}
  \caption{\label{434_over_collapse}
}
\end{figure}

\begin{figure}[htb]
\psfrag{Q}[tc][bc]{$Q$}
\psfrag{Ratio Model Contacts to Native Contacts}[cr][cr]{Total Contacts/Native Contacts}
\psfrag{0}[tc][tc]{$0$}
\psfrag{0.6}[cr][cr]{$0.6$}
\psfrag{0.8}[cr][cr]{$0.8$}
\psfrag{1}[cr][cr]{$1.0$}
\psfrag{1.2}[tc][tc]{$1.2$}
\psfrag{1.4}[tc][tc]{$1.4$}
\psfrag{0.2}[tc][tc]{$0.2$}
\psfrag{0.4}[tc][tc]{$0.4$}
\psfrag{0.6}[tc][tc]{$0.6$}
\psfrag{0.8}[tc][tc]{$0.8$}
  \centering
  \includegraphics[width=0.6\textwidth]{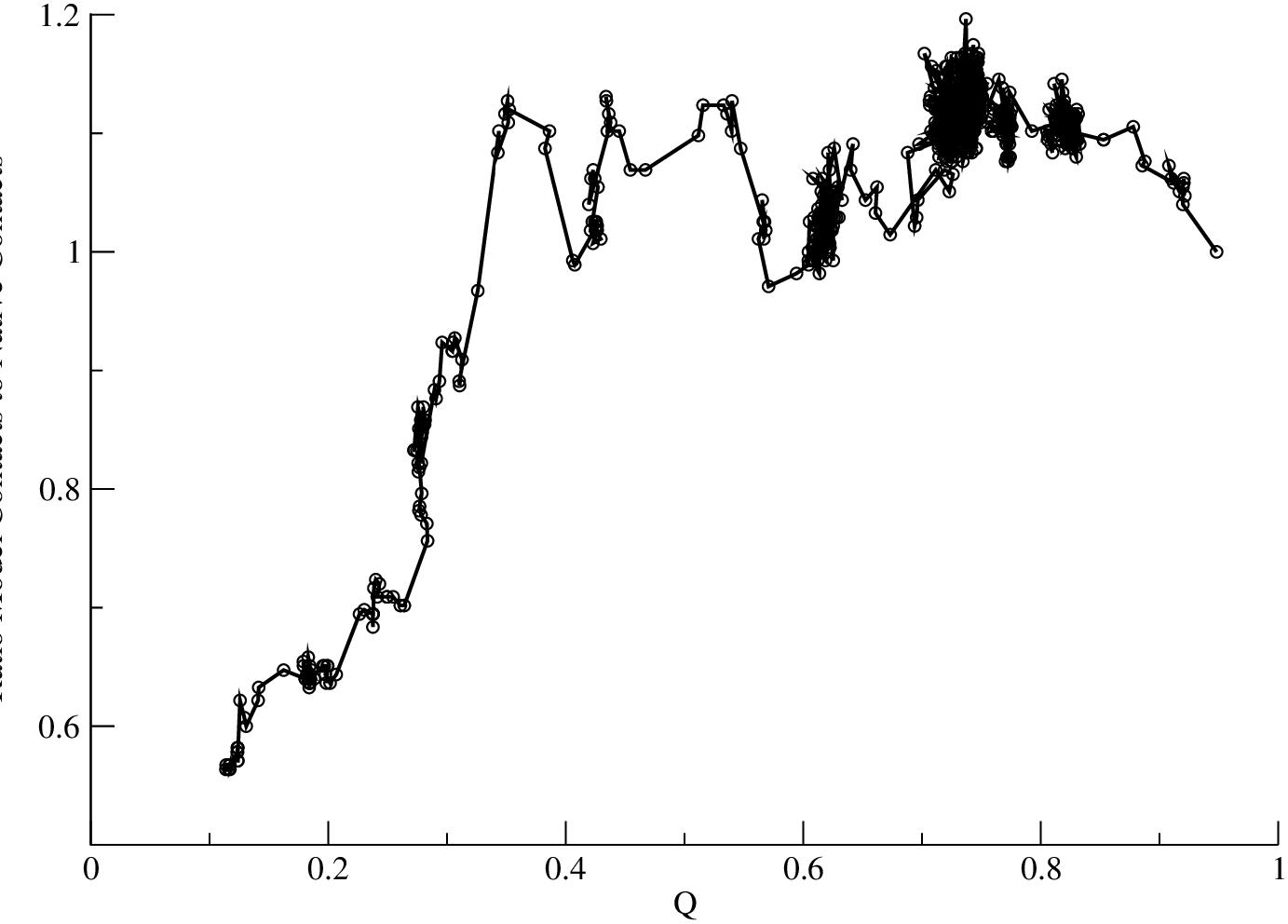}
  \caption{\label{Go_collapse} 
}
\end{figure}

\begin{figure}[htb]
\psfrag{Q}[tc][bc]{$Q$}
\psfrag{Monte Carlo Steps}[tc][bc]{Basin-Hopping Steps}
\psfrag{Ratio Model to Native Contacts}[cr][cr]{Total Contacts/Native Contacts}
\psfrag{0}[tc][tc]{$0$}
\psfrag{0.6}[cr][cr]{$0.6$}
\psfrag{0.8}[cr][cr]{$0.8$}
\psfrag{1}[cr][cr]{$1.0$}
\psfrag{1.2}[cr][cr]{$1.2$}
\psfrag{1.4}[cr][cr]{$1.4$}
\psfrag{10}[tc][tc]{$10$}
\psfrag{20}[tc][tc]{$20$}
\psfrag{30}[tc][tc]{$30$}
\psfrag{40}[tc][tc]{$40$}
\psfrag{50}[tc][tc]{$50$}
  \centering
  \includegraphics[width=0.6\textwidth]{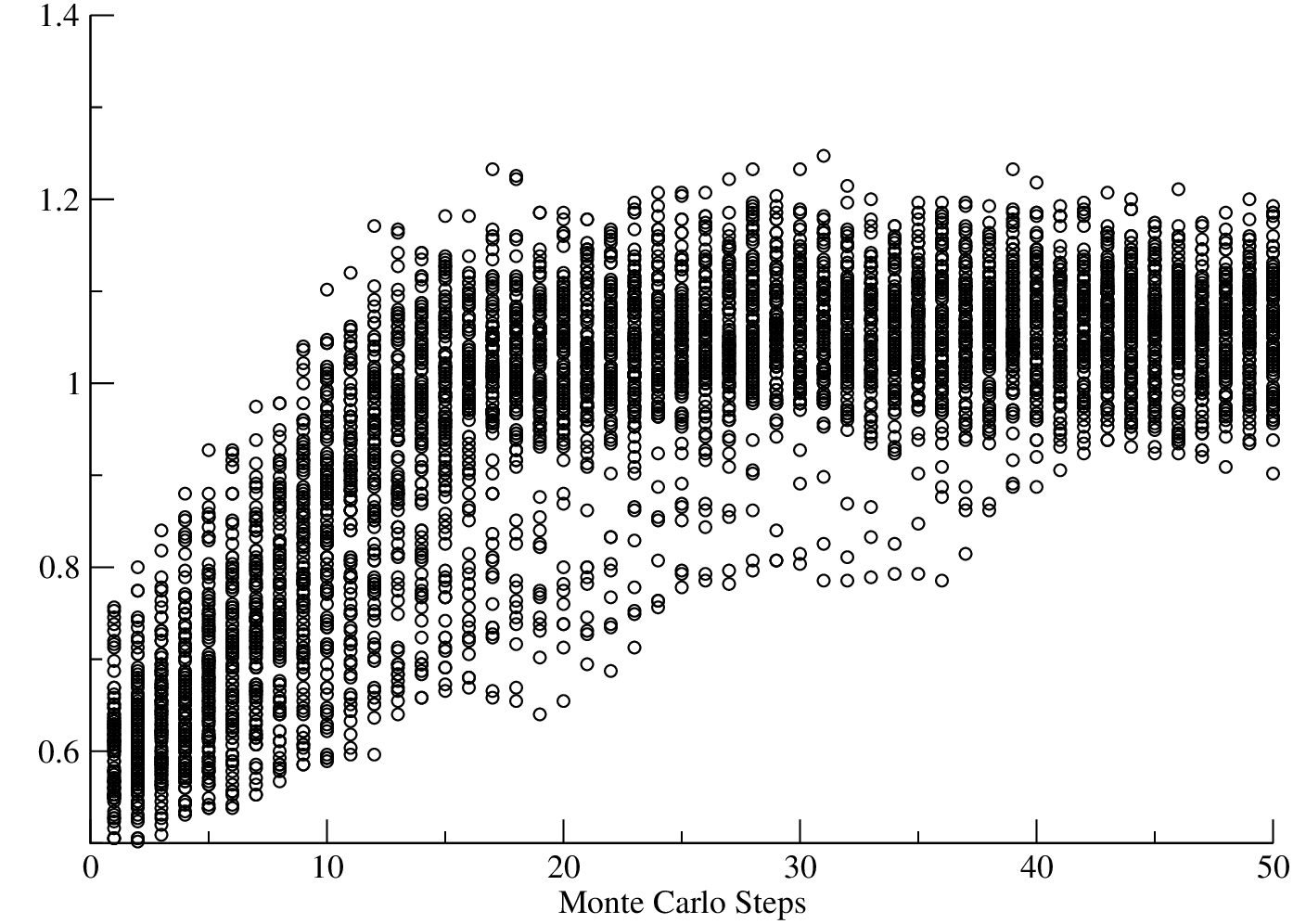}
  \caption{\label{434_fixed} 
}
\end{figure}

When we annealed the Hamiltonian using molecular dynamics we observed some 
over-collapse of the polypeptide chain, producing a smaller radius of gyration 
than the experimental structure.  In basin-hopping runs 
we also found structures exhibiting a larger number of contacts than the
experimental structure, as show in Fig.~\ref{434_over_collapse},
where a contact is defined as a C$_\alpha$-C$_\alpha$ distance of less 
than 8\,\AA. While the low-energy structures may be native-like, these 
structures were more compact than those observed experimentally.  To 
investigate this behavior, we examined the backbone and 
interaction terms of the Hamiltonian separately using the G\=o Hamiltonian 
in Eq.~(\ref{go_eq}).  Somewhat surprisingly, the G\=o model also produces 
over-collapse, as shown in Fig.~\ref{Go_collapse}.  
Hence the interaction parameters of the structure prediction Hamiltonian 
were not responsible for all of the over-collapse.  These minimal 
model-dependent frustrations were only eliminated in the final stages of 
minimisation. The most effective technique for reducing 
over-collapse was to increase the force constant and the atomic radius in 
the excluded volume terms [Eq.~(\ref{ev_eq})]. 
The barrier crossing capabilities of basin-hopping steps produce more
over-collapse than do the annealing minimisations without these 
parameter changes.  The glass-like transition seen in simulated annealing 
prevents further collapse in molecular dynamics, as the rearrangement rates 
slow down exponentially with temperature.  The improved parameter set of
Fig.~\ref{434_fixed} shows more native-like collapse, but the lowest energy
structures had $Q$ values of 0.36 and the best $Q$ value was 0.45, which 
are worse than basin-hopping simulations with the original parameters.

\section{Conclusion}
In this report we have demonstrated that minima with lower energy and 
higher quality structures can often be located for the AMH potential using 
basin-hopping global optimisation compared to annealing.  Encouragingly, the 
long range in sequence energy contributions are better minimised than 
with simulated annealing.  Umbrella sampling 
using basin-hopping can also show when the global minima are reached 
for a selected order parameter.  Previous techniques for 
reducing the energetic variance of the energy surface in simulated annealing 
are also applicable to basin-hopping.  Using basin-hopping also permits 
improvements in certain backbone terms of the Hamiltonian.  These changes 
would make the kinetics too slow in molecular dynamics annealing
runs, but larger barriers can be easily crossed using basin-hopping.

These results highlight future optimisation strategies where the 
deep non-native 
traps found by basin-hopping could be used as decoys for further parameter 
refinement, rather than the higher-lying minima obtained by quenching with 
simulated annealing.
This re-optimisation of the potential results in getting a better 
estimate for $T_{f}/T_{g}$ now possible because of the efficiency of the 
basin-hopping algorithm at identifying low energy decoys. 
Another future direction would be evaluating the equilibrium properties of 
low-lying structures identified by basin-hopping also provides a means to 
calculate free energy barriers, which would be difficult to characterise via 
conventional simulations.  

\section*{Acknowledgements}
We thank Dr Joanne Carr, and  Dr Justin Bois for helpful comments throughout 
this research.
The efforts of P.G.W. and M.C.P. are supported through the National 
Institutes of
Health Grant 5RO1GM44557.  Computing resources were supplied by the
Center for Theoretical Biological Physics through National Science
Foundation Grants PHY0216576 and PHY0225630.  M.C.P.~gratefully acknowledges 
the support by the International Institute for Complex Adaptive Matter 
(ICAM-I2CAM) NSF Grant DMR-0645461.


\begin{thebibliography}{10}
\bibitem{opt_note}
In this paper we note, the word optimisation will be used with two similar but
  distinct ways. The first use is for the calculation of the best set of
  parameters to define a minimally frustrated energy function. The second use
  is to find the global energy minima on a surface that has multiple minima of
  nearly equal energies.
\end{thebibliography}
\newpage

\section*{Figure Captions}
\begin{enumerate}
\item In the basin-hopping approach the original
potential energy surface (solid) is transformed into a set of plateaus (dashed).
The local minima are not changed, but the transition state regions are removed.
\item  The basin-hopping algorithm is defined by a few parameters that 
allow its transfer to different systems.
\item  Variation of the energy of the current minimum
as a function of $Q$ for minima encountered in the Markov chain
during a basin-hopping run using a G\=o model.
Steps that increase the energy are sometimes allowed by the Monte Carlo criterion,
which employed a temperature of 10 $k_{b}$/$\epsilon$.
\item Energy as a function of $Q$ for local minima 
of 434 repressor encountered during 100 independent basin-hopping optimisations
(top) and 20 annealing simulations (bottom).
\item 
The lowest energy structures of the training set protein, 434 repressor (top) 
and the blind prediction proteins, HDEA (bottom) identified from 100 
independent basin-hopping simulations.  Each minimum has values for energy 
illustrated by the dots and structural overlap to the 
native state $Q$ represented by lines. These minima are ordered with respect 
to their structural overlap $Q$ with the native state (Index).  The data shows 
correlations between the energy and $Q$, while the number of high quality 
structures is superior for the training protein.
\item Energy as a function of
$Q$ for the 434 repressor and cytochrome c proteins obtained in basin-hopping
calculations with the structure prediction Hamiltonian. These runs employed 
an additional umbrella potential
that constrains the simulation to different values of $Q$. 
The results for the 434 repressor are similar to the unconstrained
basin-hopping results, but the structures for cytochrome c are
$15 \epsilon$ lower in energy than those found in unconstrained
basin-hopping runs. 
\item Energies of local minima obtained using basin-hopping
with the original and a sequence-averaged Hamiltonian for two training 
proteins.  Importantly for both the top graph (434 repressor) and
the bottom graph (uteroglobin) fewer non-native states are seen with the 
sequence averaged (red) Hamiltonian when compared to standard Hamiltonian (black).
\item Results of 100 independent basin-hopping runs for the 434 repressor
using the set of backbone parameters that was optimised for 
molecular dynamics.  Structures were saved every 20 basin-hopping steps. 
The ratio of contacts to native state 
contacts shows that most of the structures are more compact than the native state.
\item A G\=o potential simulation for the 434 repressor shows a modest amount 
of over-collapse during a basin-hopping simulation, which is resolved as the
structure approaches a $Q$ value of 1.0.
\item 
Results of 100 independent basin-hopping runs for the 434 repressor using 
the set of backbone parameters that was optimised for molecular dynamics.  
Structures were saved every 20 basin-hopping steps. An altered set of 
backbone parameters produces structures that have similar 
collapse behavior when compared to the native state.
\end{enumerate} 
\newpage

\section*{Figures}

\vfill
\vfill \newpage

\vfill
\begin{figure}[htb]
\noindent
\newcounter{alines}
\begin{list}{\usecounter{alines}} 
\item \textbf{Basin Hopping Algorithm}
\item \textbf{\textit{Monte Carlo Step}} ($n$ steps)
\item \ \ random Cartesian move step with maximum distance ($d$) and temperature ($T_{bh}$)
\item \textbf{\textit{Minimisation}}
\item \ \ L-BFGS quasi-Newtonian method for optimization
\item \ \ convergence condition ($\delta E_{min}$) is RMS gradient of $10^{-3}\,\epsilon/ r$
\item \textbf{\textit{Minimisation with tight convergence}} (after $n$ steps)
\item \ \ convergence condition ($\delta E_{final}$) is RMS gradient of $10^{-5}\,\epsilon/ r$
\end{list}
\caption{\label{pseudocode}}
\end{figure}
\vfill \newpage

\vfill
\vfill \newpage

\vfill
\vfill \newpage

\vfill

\vfill \newpage

\vfill \newpage

\vfill
\vfill \newpage

\vfill \newpage

\vfill
\vfill \newpage

\vfill
\vfill \newpage

\section*{Tables}

\vfill
\begin{table}[htbp]
  \caption{\label{gmin_compare} Minima located by molecular dynamics/annealing (MD) 
and basin-hopping (BH); the first three proteins are in the training set of the Hamiltonian, 
while the results for the second three proteins are predictions.}
\bigskip
  {\centering \begin{tabular}{|c|c|c|c|c|c|c|c|c|c|} \hline
             & & MD & & & & BH &  &  &  \\ \hline 
      PDB ID & length  & Lowest $E$ & $Q$ & Highest $Q$ & $E$ & Lowest $E$ & $Q$ & Highest $Q$ & $E$  \\ \hline
1r69 & 63  & -428.92 & 0.39 & 0.53 & -307.96 & -435.82 & 0.39 & 0.52 & -408.482 \\
3icb & 75  & -536.98 & 0.47 & 0.52 & -390.54 & -546.57 & 0.40 & 0.49 & -518.92 \\
256b & 106 & -735.02 & 0.42 & 0.65 & -707.51 & -737.31 & 0.37 & 0.40 & -716.51 \\ \hline
1uzc & 69  & -457.55 & 0.36 & 0.42 & -383.08 & -458.09 & 0.37 & 0.45 & -433.41 \\
1bg8 & 76  & -469.49 & 0.25 & 0.34 & -465.19 & -468.67 & 0.36 & 0.39 & -461.50 \\
1bqv & 110 & -737.91 & 0.21 & 0.27 & -441.92 & -764.20 & 0.23 & 0.27 & -481.22 \\ \hline
    \end{tabular}\par} 
\end{table}
\vfill \newpage

\vfill
\begin{table}[htbp]
  \caption{\label{terms} Contribution of different energy terms
in local minima obtained using molecular dynamics/annealing (MD) and
basin-hopping (BH).}
\bigskip
  {\centering \begin{tabular}{|c|c|c|c|c|c|c|c|c|c|c|} \hline
      PDB &  Method & length  & Ex Vol & Rama & Short Range & Medium Range & Long Range\\ \hline
      1r69 & MD & 63 &  9.77 & $-$101.64 & $-$128.90 & $-$84.87 & $-$123.28 \\
      1r69 & BH & 63 &  2.65 & $-$91.06  & $-$125.04 & $-$84.80 & $-$137.57 \\
      3icb & MD & 75 & 11.74 & $-$127.70 & $-$177.21 & $-$90.11 & $-$153.69 \\
      3icb & BH & 75 &  4.40 & $-$115.76 & $-$178.47 & $-$83.37 & $-$173.38 \\
      1uzc & MD & 69 & 10.10 & $-$118.66 & $-$134.00 & $-$90.75 & $-$124.24 \\
      1uzc & BH & 69 & 2.22  & $-$106.20 & $-$137.95 & $-$92.40 & $-$123.77 \\
      1bg8 & MD & 76 & 11.68 & $-$136.39 & $-$173.45 & $-$94.40 & $-$76.94  \\
      1bg8 & BH & 76 & 2.72  & $-$112.13 & $-$151.95 & $-$94.23 & $-$113.09 \\ \hline
    \end{tabular}\par}
\end{table}
\vfill 

\end{document}